\documentclass[twocolumn, preprintnumbers,amsmath,amssymb,pra,aps,superscriptaddress]{revtex4-1}

\usepackage{graphicx}  
\usepackage{latexsym}

\newcommand{\um}{\mu\mathrm{m}}
\newcommand{\mm}{\mathrm{mm}}
\newcommand{\cm}{\mathrm{cm}}
\newcommand{\mK}{\mathrm{mK}}
\newcommand{\K}{\mathrm{K}}
\newcommand{\T}{\mathrm{T}}
\newcommand{\mT}{\mathrm{mT}}
\newcommand{\Qi}{Q_{\mathrm{i}}}
\newcommand{\Ohm}{\Omega}
\newcommand{\KOhm}{\mathrm{k}\Omega}
\newcommand{\Zr}{Z_{\mathrm{r}}}
\newcommand{\Bpar}{B_{||}}
\newcommand{\Bperp}{B_{\perp}}
\newcommand{\fr}{f_{\mathrm{r}}}
\newcommand{\muV}{\mu\mathrm{V}}
\newcommand{\w}{w}
\newcommand{\nm}{\mathrm{nm}}
\newcommand{\dBm}{\mathrm{dBm}}
\newcommand{\dB}{\mathrm{dB}}
\newcommand{\Pin}{P_\mathrm{in}}
\newcommand{\Tc}{T_{\mathrm{c}}}
\newcommand{\GHz}{\mathrm{GHz}}
\newcommand{\uH}{\mu\mathrm{H}}
\newcommand{\nH}{\mathrm{nH}}
\newcommand{\pH}{\mathrm{pH}}
\newcommand{\nph}{n_\mathrm{ph}}

\begin{document}

\title{High Kinetic Inductance Superconducting Nanowire Resonators \\ for Circuit QED in a Magnetic Field}

\author{N.~Samkharadze}
\affiliation{QuTech and Kavli Institute of Nanosicence, Delft University of Technology, Lorentzweg 1, 2628 CJ Delft, The Netherlands}
\author{A.~Bruno}
\affiliation{QuTech and Kavli Institute of Nanosicence, Delft University of Technology, Lorentzweg 1, 2628 CJ Delft, The Netherlands}
\author{P.~Scarlino}
\affiliation{QuTech and Kavli Institute of Nanosicence, Delft University of Technology, Lorentzweg 1, 2628 CJ Delft, The Netherlands}
\author{G.~Zheng}
\affiliation{QuTech and Kavli Institute of Nanosicence, Delft University of Technology, Lorentzweg 1, 2628 CJ Delft, The Netherlands}
\author{D.~P.~DiVincenzo}
\affiliation{JARA Institute for Quantum Information, RWTH Aachen University, D-52056 Aachen, Germany}
\author{L.~DiCarlo}
\affiliation{QuTech and Kavli Institute of Nanosicence, Delft University of Technology, Lorentzweg 1, 2628 CJ Delft, The Netherlands}
\author{L.~M.~K.~Vandersypen}
\affiliation{QuTech and Kavli Institute of Nanosicence, Delft University of Technology, Lorentzweg 1, 2628 CJ Delft, The Netherlands}

\date{\today}

\begin{abstract}

We present superconducting microwave-frequency resonators based on NbTiN nanowires. The small cross section of the nanowires minimizes vortex generation, making the resonators resilient to magnetic fields.
Measured intrinsic quality factors exceed $2\times 10^5$ in a $6~\T$ in-plane magnetic field, and $3\times 10^4$ in a $350~\mT$ perpendicular magnetic field.
Due to their high characteristic impedance, these resonators are expected to develop zero-point voltage fluctuations one order of magnitude larger than in standard coplanar waveguide resonators.
These properties make the nanowire resonators well suited for circuit QED experiments needing strong coupling to quantum systems with small electric dipole moments and requiring a magnetic field, such as electrons in single and double quantum dots.
\end{abstract}

\pacs{73.43.-f,73.63.Hs,73.43.Qt}
\keywords{}
\maketitle

Superconducting microwave-frequency resonators are widely considered essential building blocks of future quantum processors, providing a means for qubit readout and long-range interconnect in a circuit quantum electrodynamics (cQED) architecture~\cite{Blais04}. They also offer a promising interface between different types of quantum systems~\cite{Xiang13}. To reap the full benefits of cQED architectures, it is crucial to reach the strong coupling regime, wherein quantum state transfer between the qubit and the resonator is possible on a time scale shorter than the coherence time of the combined system.

Several proposals have been put forward for implementing cQED using electron spin qubits in semiconducting quantum dots~\cite{Hu12, Burkard06, Childress04, Jin12, Kloeffel13}.
Electron spins offer very long coherence times, in some case of order a second, ~\cite{Loss98, Hanson07,  Muhonen14} but convincing mechanisms for scaling in 2D are still lacking. Therefore, exploring cQED as a means for scaling is of high importance. Pioneering experiments have demonstrated coupling of superconducting cavity modes with spin and orbital degrees of freedom of the electrons~\cite{Petersson12, Frey12, Viennot15, Deng15}.

Achieving strong coupling in such hybrid systems has proved challenging due to the weak interaction between the zero-point fluctuations~(ZPF) of conventional superconducting resonators and the quantum dot electrons. Traditionally, coplanar waveguide (CPW) resonators with characteristic impedance $\Zr\sim50~\Omega$ have been used as the staple cavity in cQED. However, by increasing (decreasing)  $\Zr$, it is possible to enhance the ZPF of voltage (current), thus optimizing for electric (magnetic) dipole coupling to qubits.

Another challenge in incorporating superconducting resonators in spin- or Majorana-based systems is the typically poor performance of superconducting resonators at the magnetic fields required for the operation of such systems. Intrinsic quality factors $\Qi>10^6$ have been measured for the highest performance resonators in magnetically-shielded cQED setups~\cite{Megrant12,Bruno15}. However, strong magnetic fields induce vortices in the superconducting film, which move under the influence of microwave currents in the resonator, causing energy dissipation. A few methods have been employed to minimize vortex-induced dissipation in superconducting devices. These methods include creating artificial pinning sites and dams for the vortices~\cite{Wordenweber04, Song09, Bothner12, Bothner11}, and steering the vortices away from the areas carrying the highest currents~\cite{Wordenweber04, Vestgarden12, deGraaf12, deGraaf14}. To date, the most effective magnetic field resilience has been achieved in superconducting fractal resonators, with $\Qi\approx10^5$ in parallel magnetic field $\Bpar\approx400~\mT$~\cite{deGraaf12,deGraaf14}, and more recently, in YBCO CPW resonators with $\Qi\approx 2 \times 10^4$ at $\Bpar=7~\T$~\cite{Ghirri15}.

In this Letter, we present microwave-frequency resonators based on NbTiN nanowires, displaying magnetic field resilience and promising stronger electrical coupling.
We take advantage of the high kinetic inductance of the strongly disordered superconducting nanowires to increase $\Zr=\sqrt{\mathcal{L}/\mathcal{C}}$ and thereby also the voltage ZPF, $V^\mathrm{ZPF}_\mathrm{RMS}\propto \fr \sqrt{\Zr}$.
Here, $\fr$ is the resonance frequency and $\mathcal{L}$ $(\mathcal{C})$ is the inductance (capacitance) per unit length of the nanowire. We estimate $\Zr\approx 4~\KOhm$, nearly two orders of magnitude higher than that of CPW resonators used in typical cQED devices. The corresponding $V^\mathrm{ZPF}_\mathrm{RMS}\sim 20~\muV$~\cite{Sup}
makes these resonators well suited for coupling to systems with small electric dipole moments, such as electrons in single or double quantum dots.
Moreover, the small nanowire cross section strongly suppresses vortex generation in a magnetic field, resulting in $\Qi> 2\times10^{5}$ at up to $\Bpar=6~\T$.
We also investigate the evolution of these resonators with perpendicular magnetic field $\Bperp$, finding a clear dependence of the magnetic field resilience of $\Qi$ on the nanowire width $\w$.
The narrowest nanowires ($\w\approx100~\nm$) achieve $\Qi>3\times 10^{4}$ at $\Bperp\approx350~\mT$.

\begin{figure}
 \includegraphics[width=\columnwidth]{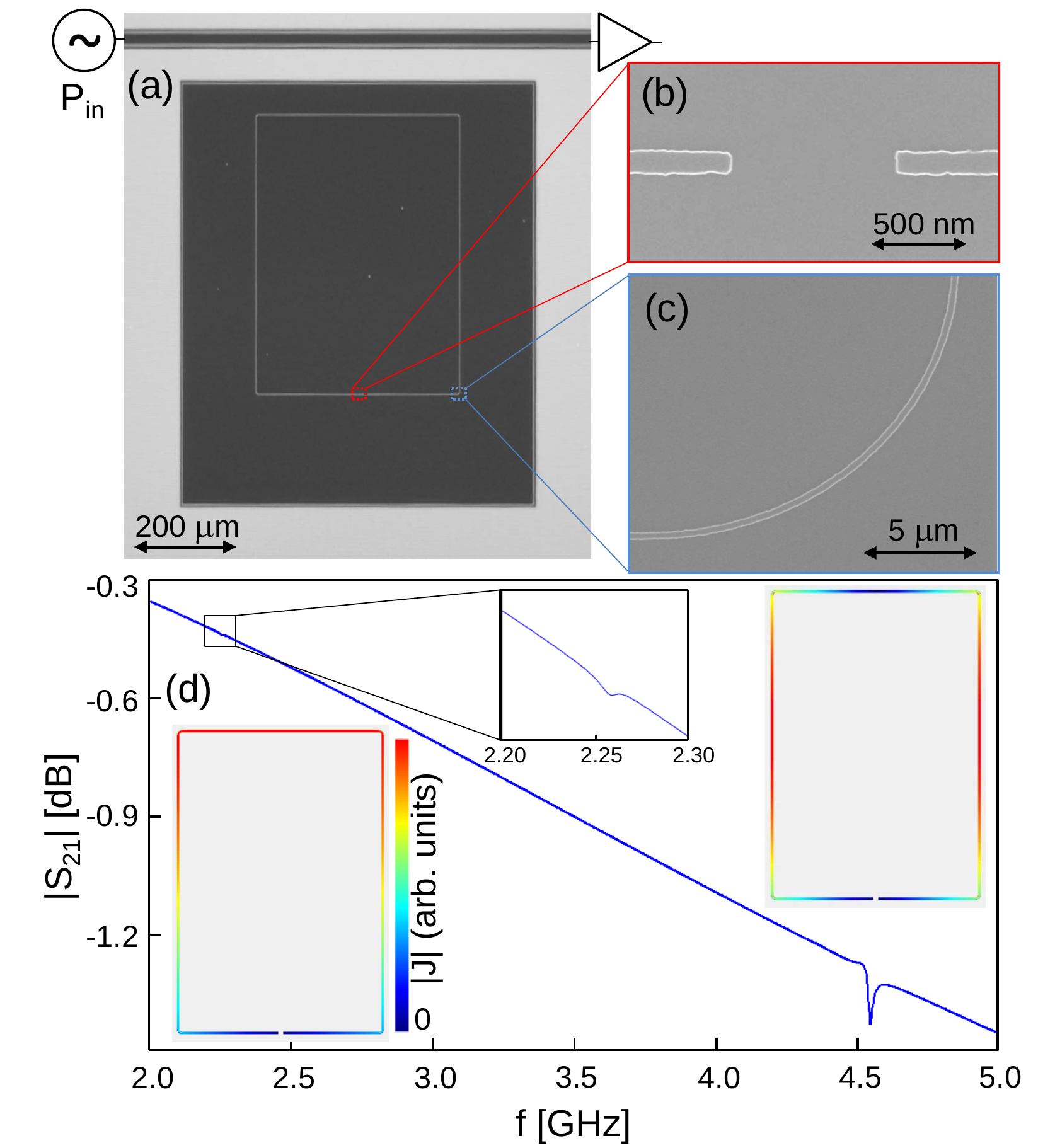}
 \caption{\label{f1}
(color online) Resonator design.
(a) Dark field optical image of a typical nanowire resonator.
(b,c) Scanning electron microscope zoom-ins of the gap (b) and the bend (c) of a typical resonator.
(d) Simulated feedline transmission for the device in (a). The insets show (absolute) current distributions along the nanowire for the fundamental and second resonance modes, as well as a zoom-in of the feedline transmission near the fundamental resonance.
}
\end{figure}

\begin{figure}
 \includegraphics[width=\columnwidth]{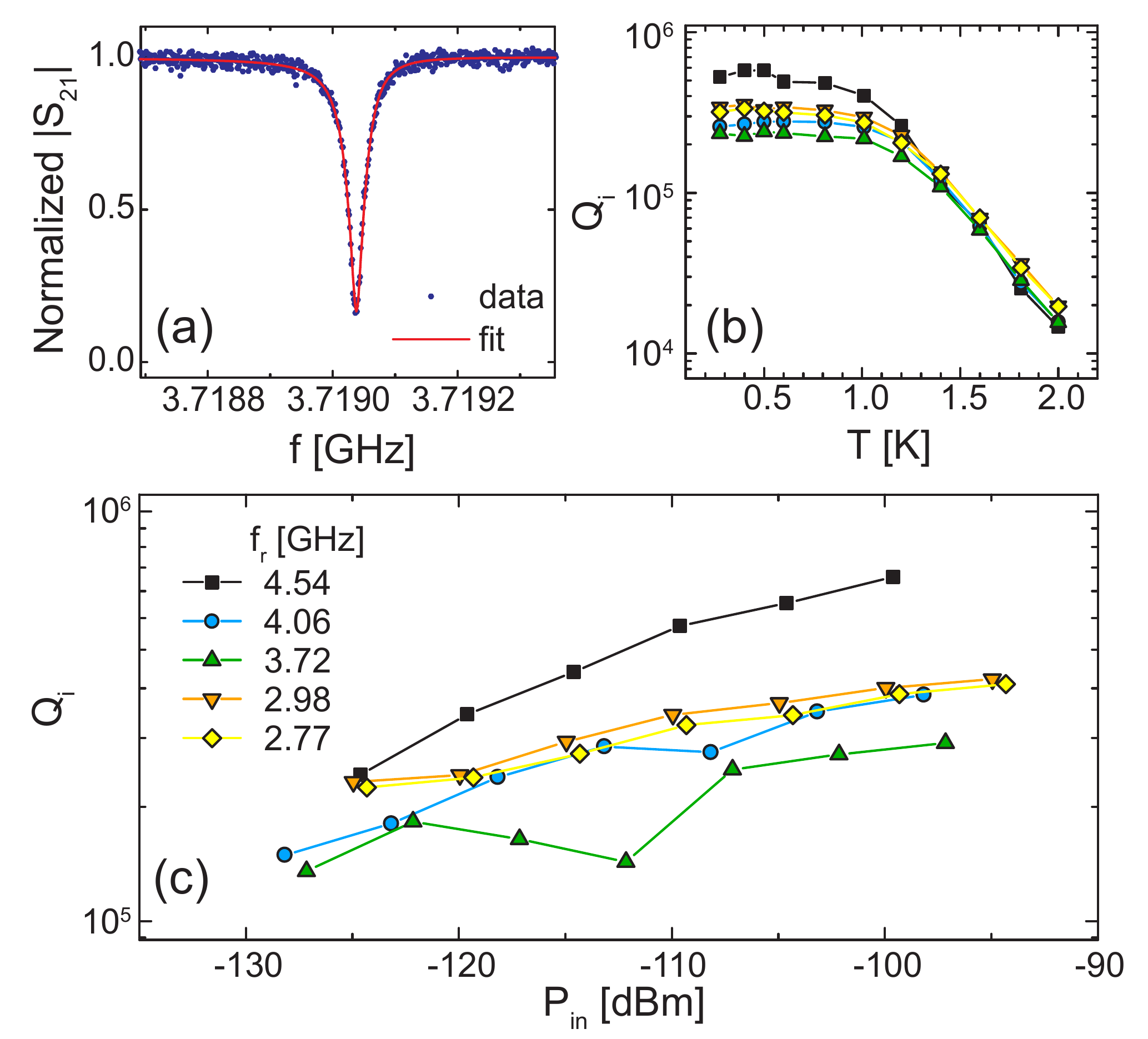}
 \caption{\label{f2}
(color online) Power and temperature dependence of intrinsic quality factors of five nanowire resonators.
 (a) Normalized absolute transmission around a typical resonance. 
     The curve is constructed from the best fit to the complex-valued feedline transmission data~\cite{Khalil12,Bruno15}.
 (b) Temperature dependence of intrinsic quality factors measured at a fixed input power $\Pin\approx-110~\dBm$. The symbols correspond to the legend in (c). Two distinct regimes are observed for $T<1~\K$ and $T>1~\K$, in which dominant loss is expected from TLS and quasiparticle dissipation, respectively.
 (c) Power dependence of intrinsic quality factors measured at $280~\mK$.  The positive slope is consistent with TLS-dominated loss.
}
\end{figure}

The resonators consist of NbTiN nanowire loops interrupted by a small gap (Fig.~1) and coupled to a common CPW feedline.
To minimize $\mathcal{C}$, the ground planes are detracted by $\sim 100~\um$ from the closest segment of the nanowire.
Figure~1(d) shows the simulated feedline transmission for the device shown in Fig.~1(a).
The ratio between the resonance frequencies of the two lowest modes extracted from the simulation is $2.01$, demonstrating that the nanowire resonators are essentially distributed resonators with a negligible direct capacitance between the nanowire ends.
In the configuration of Fig.~1(a), the coupling between the feedline and the fundamental (half-wave) mode of the resonator is inductive, which for our high impedance resonators is extremely weak.
Therefore, we here focus on the full-wave mode, leaving the discussion of the fundamental to the supplementary material~\cite{Sup}.

Device fabrication begins with the sputtering of a NbTiN film (thickness $t\sim8~\nm$) on a high-resistivity Si $\langle100\rangle$ substrate~\cite{Bruno15}. A CPW feedline and several (4 or 5) nanowire resonators are next defined in a single electron-beam lithography step, followed by reactive ion etching in a SF$_6$/He plasma. Completed devices are cooled in a $^\mathrm{3}$He refrigerator with $280~\mK$ base temperature and $70~\dB$ cold attenuation between room temperature and the feedline input. Each resonator is characterized by measuring the complex-valued feedline transmission near its resonance (Fig. 2). Fitting of the model from ref.~26 to the data allows extracting the resonance frequency and the coupling and intrinsic quality factors~\cite{Khalil12,Bruno15}.

The highly disordered nature of NbTiN and the extremely small cross-sectional area of the nanowires make the kinetic inductance the dominant contribution to the total inductance of the resonators.
From the measured critical temperature $\Tc\approx9.3~\K$ and room-temperature resistivity $\rho=200~\mu\Ohm\cm$ of the film, we estimate a sheet kinetic inductance $L_\mathrm{S} \approx 35~\pH/\Box$~\cite{Annunziata10}, close to the
value $38~\pH/\Box$ needed in a Sonnet simulation to match the resonance in Fig.~1(d) to the measurements. For a resonator of length $l=2.9~\mm$ and $\w=100~\nm$ ($2.77~\GHz$ full-wave mode), this corresponds to a total in-line inductance $\mathcal{L}l\sim1~\uH$.

Figure~2(c) shows $\Qi$ of five resonators ($w=100~\nm$) as a function of input power, $\Pin$. We find $\Qi>10^5$ at $\Pin\approx-130~\dBm$ corresponding to an average occupation of the resonator by $\langle\nph\rangle \approx10$ photons. The observed increase of $\Qi$ with $\Pin$ indicates dominant loss by coupling to spurious two-level systems (TLS) which saturate at high power~\cite{Bruno15,Gao08,OConnell08}. This conclusion is further supported by the temperature dependence of $\Qi$  at $\Pin\approx-110~\dBm$, corresponding to $\langle\nph\rangle\approx 1000$ [Figure~2(b)].
Thermally excited quasiparticles dominate loss only above $1~\K\sim \Tc/10$, consistent with previous studies of quasiparticle-induced dissipation in highly disordered thin film resonators~\cite{Coumou13}.

\begin{figure}
\includegraphics[width=\columnwidth]{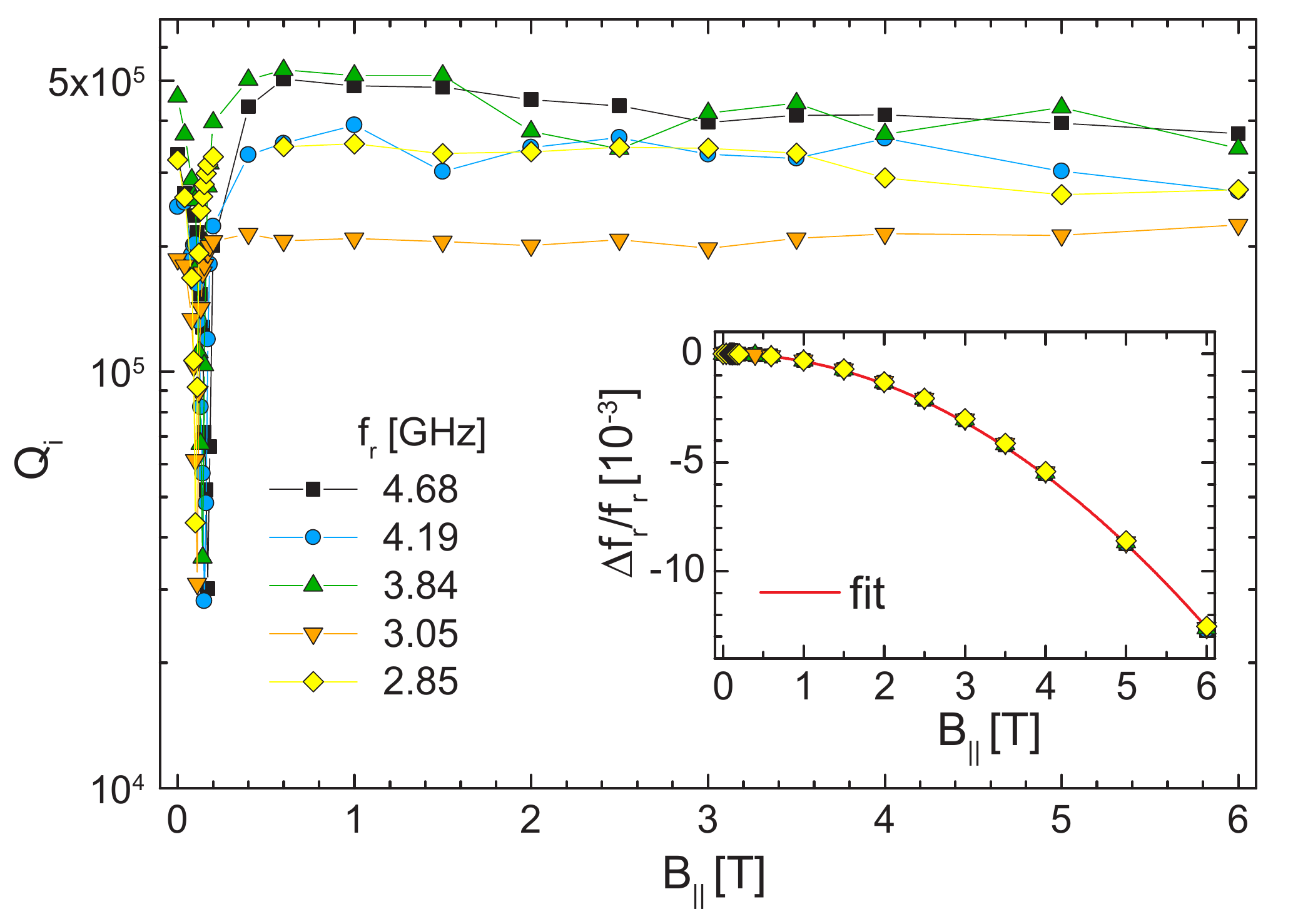}
\caption{\label{f3}
(color online) Evolution of nanowire resonator characteristics with in-plane magnetic field $\Bpar$ ($\w=100~\nm$, $T=280~\mK$, $P_\mathrm{in}\approx-110~\dBm$). The intrinsic quality factor $\Qi$ remains unaffected in the range $400~\mT \lesssim \Bpar  \leq 6~\T$. The maximum $\Bpar$ is limited by our experimental setup. We observe a sharp dip in $\Qi$ at low $\Bpar$, which we link to loading of the resonator by spurious magnetic impurities with a Land\'{e} g-factor $g\approx2$~\cite{Sup}.
(Inset) All fractional frequency shifts fit to the same simple quadratic curve.
}
\end{figure}

The resilience of the nanowire resonators to magnetic field is seen in Figure~3, which shows the typical dependence of intrinsic quality factors on the applied $\Bpar$.
Most strikingly, for $\Bpar$ between $\sim 400~\mT$ and $6~\T$, $\Qi$ is consistently above $10^5$ without sign of degradation. This field is at least one order of magnitude higher than the highest at which such $\Qi$ has been reported in earlier studies of planar superconducting resonators~\cite{deGraaf14,deGraaf12}. Moreover, we do not observe hysteretic behavior or abrupt jumps in $\fr$ with increasing $\Bpar$. These effects plague standard CPW resonators and are usually attributed to unstable magnetic flux vortices in the superconducting film~\cite{Ranjan13, Schuster10, Bothner12, deGraaf12}. These findings suggest that vortex nucleation does not take place in the nanowires. Vortices may still be created in the ground plane. However, due to the large separation between the nanowires and the ground planes, we expect only minimal current densities to be induced in the ground plane, thus weakly contributing to dissipation. Finally, we observe a sharp dip in $\Qi$ around $~100~\mT$. We link this dissipative loading to magnetic impurities with Land\'{e} g-factor $g\approx2$~\cite{Sup}. These magnetic defects likely lie in the Si substrate or at one of the interfaces.

\begin{figure}
\includegraphics[width=\columnwidth]{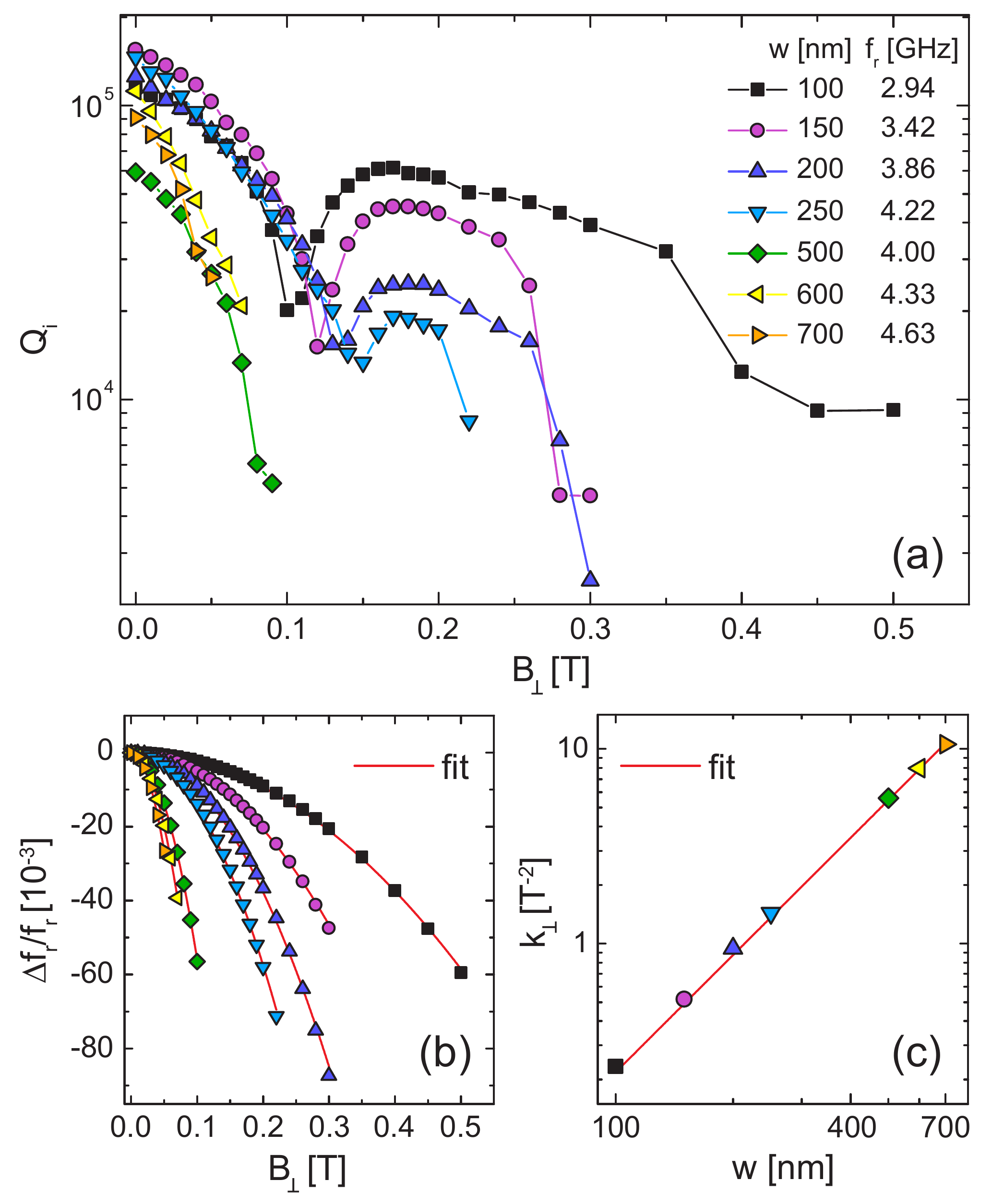}
\caption{\label{f4}
(color online) Evolution of nanowire resonator characteristics with perpendicular magnetic field, $\Bperp$. (a) $\Qi$ as a function of $\Bperp$ for various nanowire widths $\w$. The dips in $\Qi$ at low field suggest coupling to magnetic impurities, similarly to the case for $\Bpar$ in Fig.~2. The narrowest resonator retains $\Qi>3\times10^4$ up to $\Bperp=350~\mT$. (b) Fractional shift of the resonance frequencies with $\Bperp$. Same symbols as in (a). The red curves are best fits of $\Delta \fr/ \fr = -k_{\perp}(\w) \Bperp^2$ to the data.
 (c) Best-fit coefficient $k_{\perp}$ versus $\w$ and best quadratic fit.
}
\end{figure}

Further insight into the effect of magnetic field on the resonators is gained by orienting the field perpendicular to the device plane. Figure~4(a) shows the dependence of  $\Qi$ in seven nanowire resonators (widths ranging from $w=100$ to $700~\nm$) on $\Bperp$. The magnetic field resilience depends strongly on the nanowire width, and the narrowest resonators show superior performance. We observe $\Qi>3\times10^4$ for the narrowest resonator ($\w=100~\nm$) for $\Bperp\leq350~\mT$~[Fig.~4(a)]. This field range is one order of magnitude higher than the highest at which $\Qi\approx 10^4$ has been previously reported~\cite{deGraaf12}.

Turning our attention to the shift of resonance frequency induced by the magnetic field, we observe for both field orientations a quadratic shift of resonance frequency with applied field [Fig.~3 inset and Fig.~4(b)].
Fitting the fractional shifts with the expression $\Delta f_\mathrm{r}/f_\mathrm{r}=-k_{||(\perp)}B_{||(\perp)}^2$, we extract the coefficients $k_{||}$ and a width-dependent $k_{\perp}(\w)$~\cite{Healey08,Sup}.
These  coefficients reflect the increase in kinetic inductance of the superconducting nanowire due to the Cooper-pair breaking effect of the external magnetic field, and are related to the thermodynamic critical field of the superconductor: $k\propto H_\mathrm{C}^{-2}$~\cite{Healey08,Wu88}.
The penetration depth in the films $\Lambda=2\lambda^2/ t \approx 50~\um$, where $\lambda$ is the London penetration depth, is much greater than $w$. Therefore, for $B_\perp$, a Meissner state cannot be induced in the resonators and only a small fraction of the applied field is expelled.
In this case, the thermodynamic critical field  is modified as $H_\mathrm{C}\propto \w^{-1}$~\cite{Tinkham96}, which leads to $k_{\perp}(w) \propto w^2$~[Fig.~4(c)].
Furthermore, extending this geometrical scaling to the case of a parallel field yields an effective thickness of the superconductor $t_\mathrm{eff}\approx 3.5~\nm$. The reduced effective thickness of the film in the context of magnetic field expulsion is likely a combined effect of surface oxidation and the suppression of shielding currents within a coherence length from the edge.

In summary, microwave resonators based on NbTiN nanowires with extremely small cross section are highly insensitive to parallel magnetic field, with $\Qi$ remaining unaffected up to $\Bpar=6~\T$. Because of the high kinetic inductance of the nanowires, the resonators are expected to produce an order of magnitude higher vacuum voltage fluctuations compared to standard CPW resonators. Next experiments will focus on achieving strong coupling between these nanowire resonators and spin qubits in gate-defined quantum dots, which have small electric dipole moments and require a magnetic field.

We thank L.~P.~Kouwenhoven and his team for sputtering of NbTiN thin films, and G.~de~Lange, T.~M.~Klapwijk, and A.~Wallraff for fruitful discussions.
We acknowledge funding by an ERC Synergy Grant, the Dutch Organization for Fundamental Research on Matter (FOM), the Army Research Office (W911NF-12-0607), and Microsoft Corporation Station Q.

\bibliographystyle{apsrev4-1}

\newpage
\setcounter{figure}{0}
\renewcommand{\thefigure}{S\arabic{figure}}

\onecolumngrid
\section{Supplementary Information}

\title{Supplement to: High Kinetic Inductance Superconducting Nanowire Resonators \\ for Circuit QED in a Magnetic Field}

\author{N.~Samkharadze}
\affiliation{QuTech and Kavli Institute of Nanosicence, Delft University of Technology, Lorentzweg 1, 2628 CJ Delft, The Netherlands}
\author{A.~Bruno}
\affiliation{QuTech and Kavli Institute of Nanosicence, Delft University of Technology, Lorentzweg 1, 2628 CJ Delft, The Netherlands}
\author{P.~Scarlino}
\affiliation{QuTech and Kavli Institute of Nanosicence, Delft University of Technology, Lorentzweg 1, 2628 CJ Delft, The Netherlands}
\author{G.~Zheng}
\affiliation{QuTech and Kavli Institute of Nanosicence, Delft University of Technology, Lorentzweg 1, 2628 CJ Delft, The Netherlands}
\author{D.~P.~DiVincenzo}
\affiliation{JARA Institute for Quantum Information, RWTH Aachen University, D-52056 Aachen, Germany}
\author{L.~DiCarlo}
\affiliation{QuTech and Kavli Institute of Nanosicence, Delft University of Technology, Lorentzweg 1, 2628 CJ Delft, The Netherlands}
\author{L.~M.~K.~Vandersypen}
\affiliation{QuTech and Kavli Institute of Nanosicence, Delft University of Technology, Lorentzweg 1, 2628 CJ Delft, The Netherlands}

\date{\today}

\maketitle

This supplement provides calculations and additional  data sets supporting the claims made in the main text. First, we show the data on the fundamental mode of the nanowire resonator, and derive the magnitude of zero-point voltage fluctuations on the ends of the resonator in the fundamental mode.
Next, we provide additional data showing the coupling of the resonators to the magnetic impurities, and the performance of resonators with different widths in parallel magnetic field.
Finally, we present the reasoning used to explain the scaling of frequency shifts with the applied perpendicular magnetic field and with the nanowire width.

\subsection{Study of the fundamental mode}

\begin{figure}[!htb]
	\includegraphics[width=0.6\columnwidth]{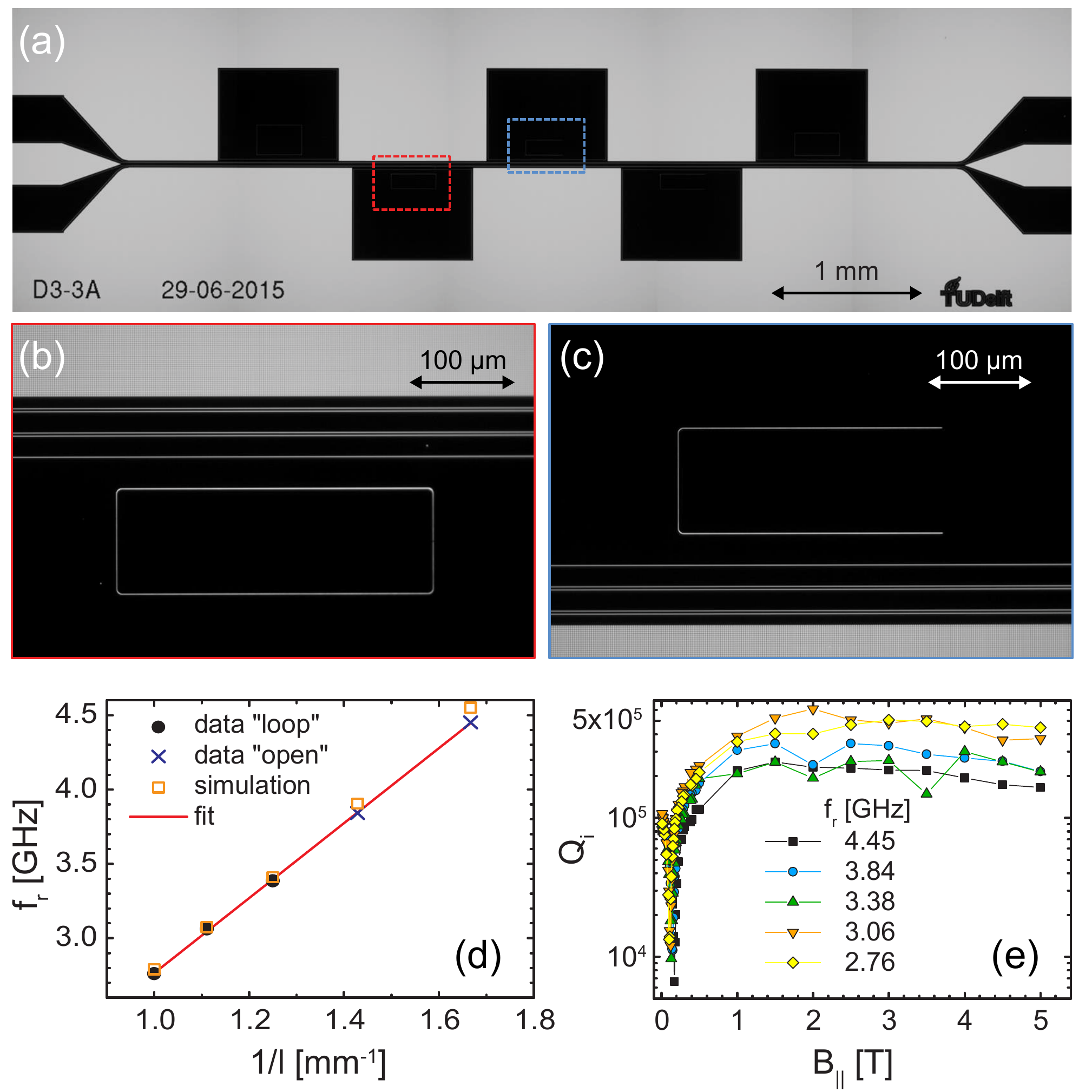}
	\caption{
(a) Dark field micrograph of a typical device with 5 nanowire resonators.
(b, c) Expanded regions from (a) showing two nanowire resonators with "loop" (b) and "open" (c) geometries, respectively. Unlike the resonators shown in the main text, the fundamental modes of these resonators couple mainly capacitively to the feedline.
(d) Linear dependence of the fundamental frequency of nanowire resonators ($\w=100~\nm$) on the inverse of their length, $l$. Frequencies are independent of how the nanowire winds.
(e) Evolution of the intrinsic quality factor of the fundamental modes with $\Bpar$.
}
	\label{fig:S1}
\end{figure}

For the fundamental mode of the nanowire resonator, the voltages at the two ends of the nanowire oscillate out of phase.
In order to increase the coupling of the fundamental mode to the feedline, we rotate the resonator by $90^{\circ}$ to enhance the capacitive coupling component ~[Fig.~S1(a-c)].
Figure~S1(d) shows the dependence of the fundamental frequencies of five nanowire resonators ($\w=100~\nm$) on the inverse of their total length, $l$.
The linear dependence of the resonance frequencies on $1/l$ is consistent with the nanowire resonators being distributed half-wave resonators with negligible direct capacitance between the nanowire ends.
To further test this hypothesis, we fabricated two of the five resonators in an "open" geometry [Fig.~S1(c), crosses in Fig.~S1(d)] with the ends facing outwards.
We find the resonance frequencies to be independent of the nanowire winding.

The NbTiN film used in the fabrication of the new sample, was deposited a few months after the film used in the main text. Based on Sonnet simulations of the resonance frequencies~[as in Fig.~S1(d)], we estimate $L_\mathrm{S}\approx75~\pH/\Box$ for the new film. This value is a factor of 2 higher than that of the film used in the main text, suggesting higher degree of disorder.

Figure~S1(e) shows the performance of these resonators as a function of parallel magnetic field at $\Pin\approx -110~\dBm$. At $\Bpar=0$, the intrinsic quality factors are somewhat lower than those shown in Figs.~2~and~3. However, as the magnetic field is applied, the quality factors are enhanced and by $\Bpar\sim 2~\T$ become comparable to those reported in the main text.

\newpage
\subsection{Zero-point voltage fluctuations at the ends of the nanowire resonator}

Figure S1 demonstrates that the nanowire resonator acts as a distributed half-wavelength resonator. Thus, in the lowest mode, current distribution on the  resonator can be expressed as

\begin{equation}
I(x,t)=I_{0} \mathrm{sin} \left( \frac{x}{l}\pi \right) \mathrm{sin} \left( \omega t \right) \mathrm{,}
\end{equation}
where $l$ is the length of the wire. The voltage difference over a small wire segment of length $dx$ a distance $x$ from the end is given by
\begin{equation}
dV_{x} = \mathcal{L} dx  \frac{\partial I(x,t)}{\partial t} \mathrm{,}
\end{equation}
where $\mathcal{L}$ is inductance per unit length. Plugging in the expression for $I(x,t)$ from Eq.~(S1) into Eq.~(S2) gives
\begin{equation}
dV_{x}=\mathcal{L} dx I_{0}\mathrm{sin} \left( \frac{x}{l}\pi \right) \omega \mathrm{cos}(\omega t)  \mathrm{.}
\end{equation}
Integrating the voltage from Eq.~(S3) over the length of the wire, we arrive at the expression for the voltage difference between the two ends of the resonator:
\begin{align}
\nonumber
\Delta V &= \mathcal{L} I_{0} \omega \mathrm{cos}(\omega t) \int^l_0 \mathrm{sin} \left( \frac{x}{l}\pi \right) \,dx \\
\nonumber
&= \mathcal{L} I_{0} \omega \mathrm{cos}(\omega t) \frac{l}{\pi} \int^{\pi}_0 \mathrm{sin} \left( \frac{x}{l}\pi \right) \,d\left( \frac{x}{l}\pi \right) \\
&=\mathcal{L} I_{0} \omega \mathrm{cos}(\omega t) \frac{2l}{\pi} \mathrm{.}
\end{align}

Next, we estimate the amplitude of the ZPF current $I_{0}$. The average energy stored in the inductance equals half of the zero-point energy:

\begin{align}
\nonumber
\frac{1}{4}\hbar\omega &= 1/2 \frac{1}{T} \int^T_0 \int^l_0 \mathcal{L} I^{2} \,dx \,dt \\
\nonumber
&= 1/2 \frac{I_{0}^{2}}{T} \mathcal{L} \int^T_0 \mathrm{sin}^2 (\omega t) \,dt \int^l_0 \mathrm{sin}^2  \left( \frac{x}{l}\pi \right) =1/8 I_{0}^{2} \mathcal{L} l  \mathrm{.}
\end{align}
Therefore,
\begin{align}
I_{0} &= \sqrt{\frac{2\hbar \omega}{\mathcal{L} l}} \mathrm{.}
\end{align}

Inserting the expression for $I_0$ from Eq.~(S5) into Eq.~(S4), we get the final expression for the voltage ZPF between two ends of the resonator:
\begin{align}
\nonumber
\Delta V &= \frac{2L}{\pi} \sqrt{\frac{2\hbar \omega}{L}} \omega \mathrm{cos}(\omega t) \mathrm{.}
\end{align}
Here $L=\mathcal{L} l$ is the total inductance of the resonator.

For the $4.45~\GHz$ resonator in Fig.~S1: $L_{S}=75~\pH/\Box$, $l=600~\um$, and $w=100~\nm$. From these values we calculate $L\approx 450~\nH$, 
and $\Delta V_\mathrm{RMS} \approx 20~\muV$.

\newpage
\subsection{Coupling to magnetic impurities}

\begin{figure}[!htb]
	\includegraphics[width=0.6\columnwidth]{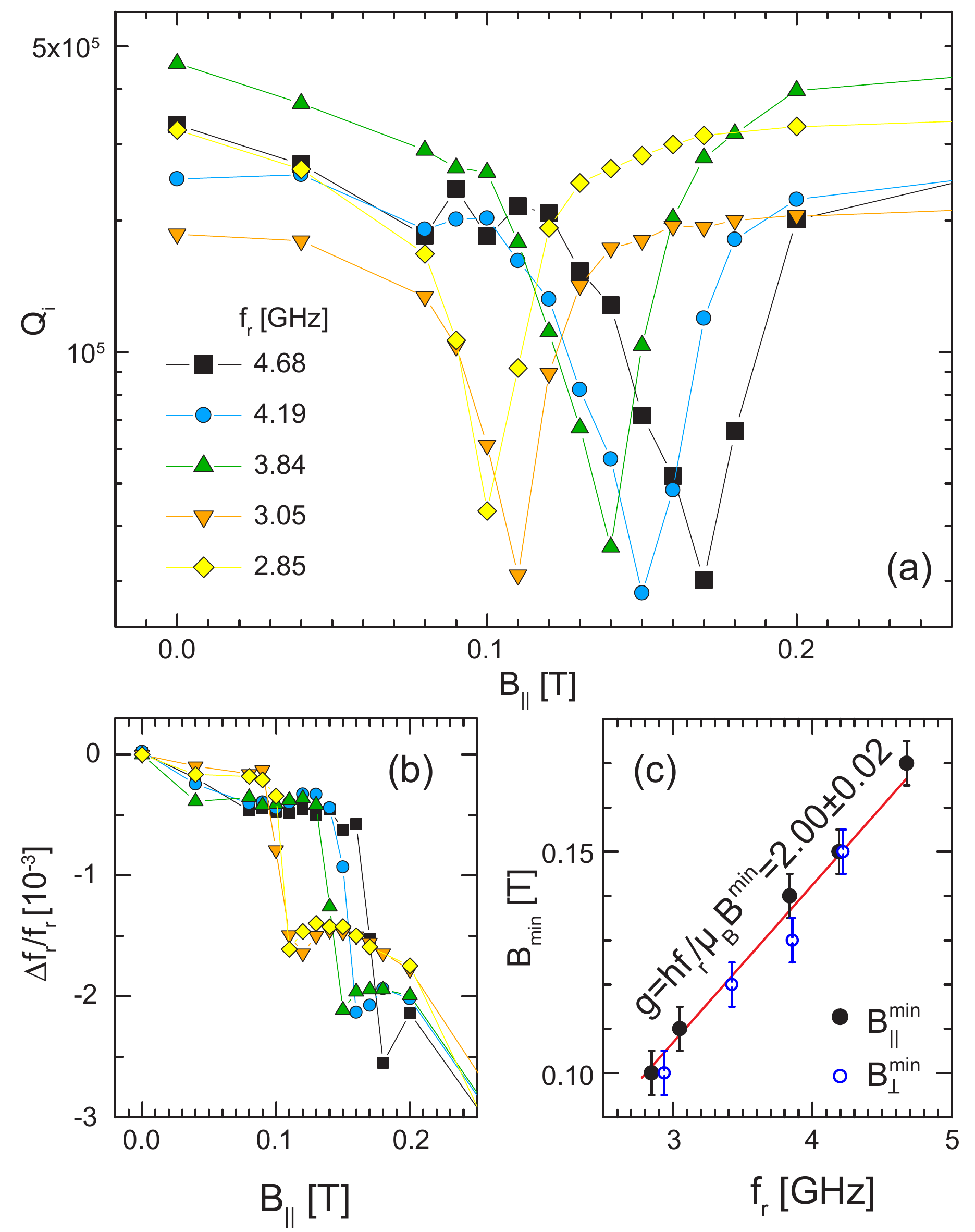}
	\caption[width=\textwidth]{Signatures of electron spin resonance near the Zeeman field for five nanowire resonators.
(a,b) Data from Fig.~3, expanded for clarity around $100~\mT$.
The minima of the quality factors of the resonators occur at different values of magnetic field.
(c) Dependence of the magnetic field positions of quality factor minima on the resonator frequencies.
Black points correspond to $\Bpar$ measurements~[Fig.~S2(a), Fig.~3], and blue points to $\Bperp$ measurements ~[Fig.~4(a)]. The straight line is the best fit to the data.}
	\label{fig:fridges}
\end{figure}

Figures 3 and 4 show sharp dips in the quality factors of the resonators  around $B_{||,\perp}=100\,$mT. Upon closer inspection, it is evident that the magnetic field values, at which these dips occur, scale with the frequency of the resonators~[Fig.~S2(a, c)]. This suggests that the resonators couple with  magnetic impurities in the silicon substrate or at one of the interfaces. Moreover, magnetic field dependence of the frequency shifts of the resonators shows an incipient avoided crossing~[Fig.~S2(b)]. Fitting the frequency dependence of the  magnetic field positions of the quality factor minima with the condition for spin resonance $hf_r=g\mu_\mathrm{B}B$, we extract the value for Land\'e $g$-factor: $g=2.00\pm 0.02$~[Fig.~S2(c)].

\newpage
\subsection{Resonator width dependence of the performance in parallel magnetic field}

\begin{figure}[!htb]
	\includegraphics[width=0.6\columnwidth]{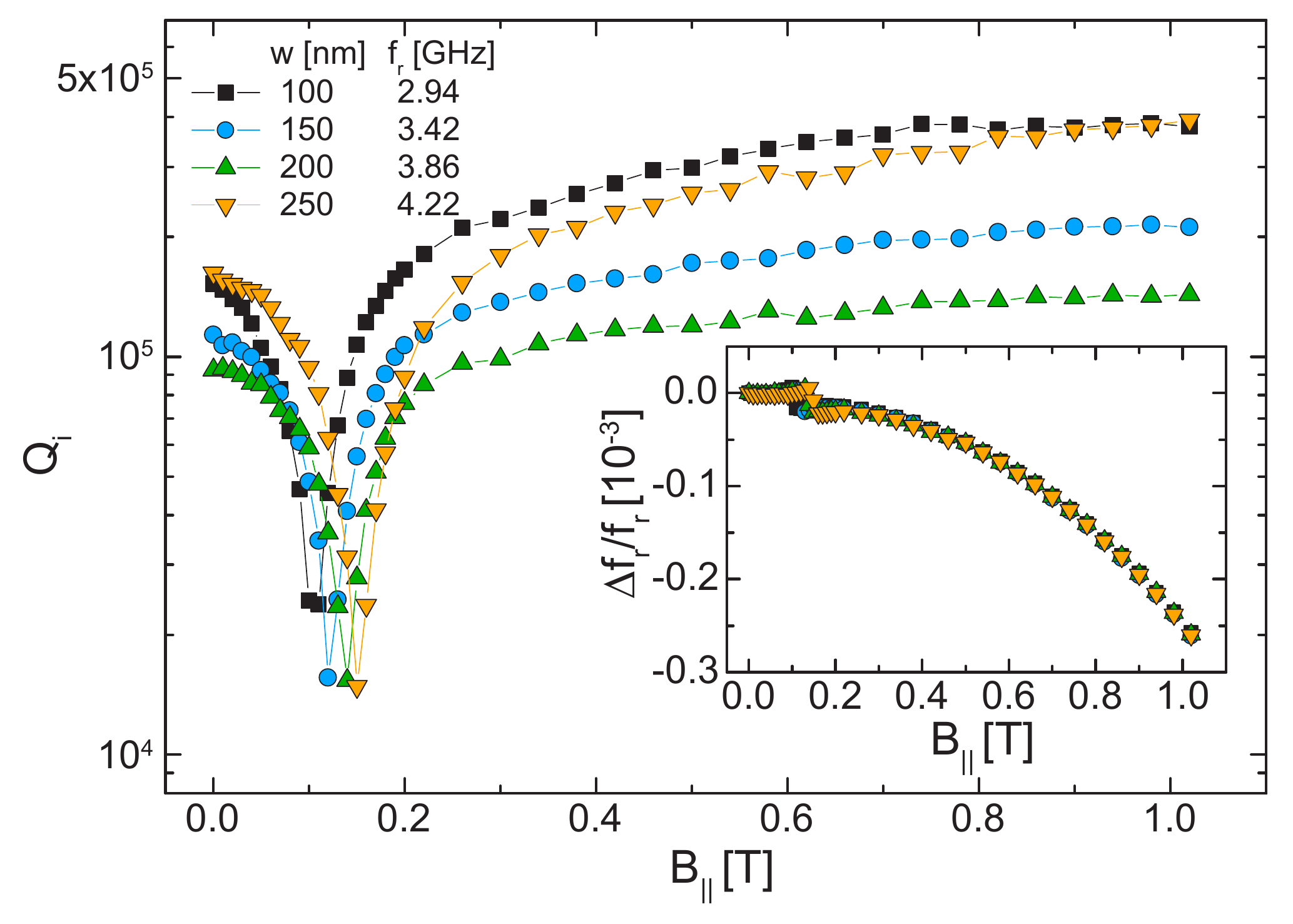}
	\caption{$\Bpar$ evolution of $\Qi$ in the four narrowest nanowire resonators shown on Fig.~4.
(Inset) Fractional shifts of the four resonance frequencies as a function of the applied field. Symbols correspond to those in Fig.~4.}
	\label{Fig:fitting}
\end{figure}

Figure S3 shows the $\Bpar$ evolution of $\Qi$ and $\fr$ for the four resonators from Fig.~4 with narrowest nanowires. The fractional frequency shifts for all resonators collapse onto a single curve, demonstrating that the contribution from any out-of plane component due to field misalignment is negligible.

\newpage
\subsection{Frequency shift in perpendicular field}
Figure 4(b, c) demonstrated the scaling of fractional frequency shift with the square of $\Bperp$ and nanowire width $\w$.
Taking into account that the dominant contribution to the nanowire inductance is kinetic: $\fr \propto L_\mathrm{k}^{-1/2}$, where $L_\mathrm{k}$ is the kinetic inductance of the resonator.
Further, for $T\ll\Tc$: $L_\mathrm{k} \propto \Tc^{-1}$~\cite{Annunziata10_1}, and for small changes in frequency: $\Delta \fr/\fr= -\frac{1}{2} \Delta L_\mathrm{k}/L_\mathrm{k} = \frac{1}{2} \Delta T_\mathrm{C}/T_\mathrm{C}$.

The applied magnetic field splits the time-reversal degeneracy of the paired electrons, giving rise to an effective depairing energy $2\alpha$~\cite{Tinkham96_1}.
In the dirty limit and for small $\alpha$, the change in $\Tc$ due to this pair-breaking effect is linear in $\alpha$: $k_\mathrm{B} \Delta \Tc=-\frac{\pi}{4}\alpha$.
Keeping in mind that the effective perpendicular penetration depth is much larger than the nanowire width, we  make use of the  expression for $\alpha$ valid in the  "thin film in parallel field" approximation, $\alpha=\frac{1}{6} \frac{D e^2 B_{\perp}^2 w^2}{\hbar}$, where $D$ is the electronic diffusion constant~\cite{Tinkham96_1}. Thus, we recover the experimentally observed scaling $\Delta \fr/\fr =-\frac{\pi}{48} \frac{De^2}{\hbar k_\mathrm{B} \Tc}\Bperp^2 \w^2$, and extract the diffusion constant $D\approx 2 \, \mathrm{cm}^2 \mathrm{s}^{-1}$.
This value is consistent with an earlier estimate~\cite{Tong01} of the electronic diffusion constant in NbTiN thin films.

\bibliographystyle{apsrev4-1}

\end{document}